# Decoupled photoelectrochemical water splitting system for centralized hydrogen production

Avigail Landman[1,5], Rawan Halabi[2,5], Paula Dias[3], Hen Dotan[2], Alexander Mehlman[2], Gennady E. Shter[4], Manar Halabi[4], Omayer Naserladeen[4], Adélio Mendes[3], Gideon S. Grader[1,4,*] and Avner Rothschild[1,2,6,*]

[1]The Nancy and Stephen Grand Technion Energy Program (GTEP), Technion – Israel Institute of Technology, Technion City, Haifa 3200003, Israel

[2]Department of Materials Science and Engineering, Technion – Israel Institute of Technology, Technion City, Haifa 3200003, Israel

[3]LEPABE – Faculty of Engineering, University of Porto, Rua Dr. Roberto Frias, 4200-465 Porto, Portugal

[4]Department of Chemical Engineering, Technion – Israel Institute of Technology, Technion City, Haifa 3200003, Israel

[5] *These authors contributed equally to this work*

[6] *Lead contact*

*\* Corresponding authors. Email addresses: [avnerrot@technion.ac.il](mailto:avnerrot@technion.ac.il) (A. R.); [grader@technion.ac.il](mailto:grader@technion.ac.il) (G. S. G.)*

**Summary**

Photoelectrochemical (PEC) water splitting offers an elegant approach for solar energy conversion into hydrogen fuel. Large-scale hydrogen production requires stable and efficient photoelectrodes and scalable PEC cells that are fitted for safe and cost-effective operation. One of the greatest challenges is the collection of hydrogen gas from millions of PEC cells distributed in the solar field.  In this work, a separate-cell PEC system with decoupled hydrogen and oxygen cells was designed for centralized hydrogen production, using 100 cm$^2$ hematite ($\alpha$-Fe$_2$O$_3$) photoanodes and nickel hydroxide (Ni(OH)$_2$) / oxyhydroxide (NiOOH) electrodes as redox mediators. The operating conditions of the system components and their configuration were optimized for daily cycles, and ten 8.3 h cycles were carried out under solar simulated illumination without additional bias at an average short-circuit current of 55.2 mA. These results demonstrate successful operation of a decoupled PEC water splitting system with separate hydrogen and oxygen cells.

## Introduction

Widespread use of solar power requires an efficient strategy for its harvesting, conversion and storage.[1] The production of chemical fuels is an attractive route for storing energy from sunlight. Particularly, hydrogen generation by solar water splitting represents a promising path towards the realization of a sustainable carbon-neutral society.[2,3] Hydrogen is an energy dense, storable and transportable fuel, which can be used as a feedstock for fuel cells and converted into electricity for stationary or mobile applications. It can also be used for the synthesis of other fuels and in the chemical industry.[4]

Photovoltaic-powered water splitting (PV-electrolysis) systems that couple commercially-available photovoltaic and water electrolysis technologies have already been demonstrated in several pilot plants and hydrogen refueling stations.[5–7] The highest reported solar to hydrogen (STH) conversion efficiency for such a system composed of polymer electrolyte membrane (PEM) electrolyzers powered by an InGaP/GaAs/GaInNAsSb triple-junction solar cell was 30%, tested over 48 h.[8] Despite the high efficiency, the device's complexity and cost makes its upscale potential impractical. PV-electrolysis systems comprising conventional Si PV modules and alkaline electrolyzers typically achieve STH efficiency of less than 10%.[9] Inspired by natural photosynthesis,[10] photoelectrochemical (PEC) water splitting that combines light harvesting and electrochemical conversion of electrical power to chemical energy stored in hydrogen bonds, wherein both functions are carried out concurrently at the solid/liquid interface between a semiconductor photoelectrode and water,[11–13] aims at providing a competitive solution for solar energy conversion and storage.[14,15]

Critical challenges towards economically viable PEC hydrogen production technologies include: (i) the development of stable and efficient photoelectrodes made of earth-abundant elements; and (ii) the development of PEC cell architectures suitable for large-scale hydrogen production. These challenges have yet to be overcome and both are still in the research and development stage.

Since the pioneering work of Fujishima and Honda in 1972,[11] the identification of a single semiconductor material that can both develop sufficient photovoltage for overall water splitting (1.5 - 1.8 V) and harvest a significant fraction of the solar spectrum remains challenging.[16,17] Hematite ($\alpha$-$Fe_2O_3$) emerged as one of the most promising photoanodes due to its abundance, stability and suitable bandgap energy.[18] In spite of significant advancements in this field,[19–21] hematite photoanodes have yet to reach their full theoretical capabilities, and research is still needed to improve their performance and achieve higher photocurrents at lower potentials. Specifically, further research is required to understand the underlying causes of physical and chemical phenomena that limit the photocurrent and photovoltage in hematite.[22–24] Hematite photovoltage is insufficient to drive water splitting without additional bias.[25] This shortcoming is common to most of the photoelectrode materials known to date. It is commonly addressed by tandem connection of multi-photoabsorbers, each absorbing complementary fractions of the solar spectrum, that provide sufficient photovoltage to drive overall water splitting. A photoelectrode-photovoltaic (PEC-PV) tandem configuration is envisaged as the most promising approach, with the potential to surpass the efficiency of PV-electrolysis systems using the same PV cells.[15] In an ideal PEC-PV tandem arrangement, a wider-bandgap and stable metal oxide photoelectrode (1.8 – 2.4 eV) should be placed on top of a smaller-bandgap PV cell (1.0 – 1.5 eV) in a series connection.[26,27] PEC-PV cells with hematite photoanodes have been reported, with solar to hydrogen (STH) conversion efficiency up to 3.4% for small-area devices.[28–32]

Over the past few years, significant advances have been made in the development of materials for individual components as well as in the design of devices at the lab-scale level.[33] However, large-scale PEC cell design is a crucial driver to develop competitive PEC water splitting technology, requiring innovative engineering solutions.[34,35] Thus far, only a handful of studies proposed scalable PEC-PV tandem devices.[31,36–39] The typical PEC cell design, presented in *Figure 1*A, is based on the conventional water electrolysis cell wherein the water oxidation and reduction reactions are tightly coupled, proceeding simultaneously in the same cell. The cell comprises an hermetically sealed body with both electrodes (anode and cathode) immersed in the electrolyte, placed back-to-back or side-by-side and separated by a membrane or diaphragm to avoid $O_2/H_2$ cross-over; PortoCell[38] and CoolPEC[31] are two examples of this architecture. Although this single-cell architecture is well-suited for separation and collection of hydrogen on a small scale, the large number of PEC cells that are needed in industrial applications due to the low power density of the sunlight gives rise to technical and economic challenges on a large scale. The construction material and separator costs, as well as the hydrogen gas piping manifold construction and maintenance costs renders large-scale PEC water splitting in this single-cell architecture economically questionable.[40]

➔Figure 1

A membrane-free decoupled water splitting approach wherein hydrogen and oxygen are produced in two separate cells that are connected to each other only electrically, as illustrated in *Figure 1*B, was proposed recently.[40,41] This approach offers an alternative solution for the challenge of $H_2/O_2$ gas separation and hydrogen collection from distributed PEC cells, as well as other advantages of decoupled water splitting, as reported elsewhere.[42–44] The ion exchange between the hydrogen and oxygen cells is mediated by a pair of nickel hydroxide-based auxiliary electrodes of the sort used as the positive electrode in rechargeable alkaline batteries.[45] The oxygen cell contains the photoanode, which can be connected to a PV cell that provides the necessary bias for unassisted solar water splitting as in PEC-PV tandem cells, while the hydrogen cell is an electrolytic cell. The oxygen cell carries out the water (photo)oxidation / oxygen evolution reaction, while the hydrogen cell carries out the water reduction / hydrogen evolution reaction. The driving force for both reactions is provided by the electromotive force / photovoltage generated at the photoanode and the PV cell connected to it in a tandem configuration. Thus, only the oxygen cell must be placed at the solar field and the oxygen evolved at the photoanode can be released to the atmosphere. Meanwhile, the hydrogen can be produced in a central generator placed at the edge of the solar field, as illustrated in *Figure 2*. This separate-cell arrangement eliminates the need to hermetically seal the solar panels, as well as the need to collect the hydrogen gas from each and every one of the PEC cells that are distributed in the solar field, as in the conventional PEC cell configuration (*Figure 1*A). Thus, instead of collecting hydrogen gas from all the PEC cells and transporting it through a multiplexed gas manifold to a central storing and distribution vessel, the hydrogen is produced in a central generator in the proposed configuration. It is noted that the auxiliary electrodes in the hydrogen and oxygen cells have finite capacity, and therefore they must be swapped between the respective cells when they are charged or discharged to their full operational capacity. Therefore, instead of collecting and transporting hydrogen gas from distributed PEC cells in the solar field in the conventional PEC cell design, nickel-based solid electrodes must be collected and swapped periodically between the distributed PEC oxygen cells and the central hydrogen generator in our proposed separate-cell configuration. This concept was demonstrated in a lab-scale electrolytic configuration with nickel anode and cathode, powered by a PV mini-module, yielding unassisted solar water splitting with an STH efficiency of 7.5%.[40] However, it has never been demonstrated in a photoelectrochemical water splitting device as illustrated in *Figure 1*B.

➔Figure 2

The present work complements our earlier study in which the conceptual idea of cell separation was proposed and demonstrated in a purely electrolytic setup,[40] demonstrating a benchtop-scale separate-cell tandem PEC-PV device for decoupled photoelectrochemical water splitting in separate oxygen and hydrogen cells. It addresses the challenges of designing, building and optimizing the device for assessing large-scale hydrogen generation. The oxygen cell contains two 100 cm$^2$ back-to-back hematite photoanodes, placed in tandem with Si PV mini-modules that provide the necessary bias for driving unassisted solar water splitting. The hydrogen cell contains the cathode, and it is physically separated from the oxygen cell. Battery-grade nickel hydroxide electrodes are placed in both cells to mediate the ion (OH$^-$) exchange between the cathode and anode. Successful operation of this prototype system was also demonstrated in outdoor conditions with natural sunlight.

**System components**

**Hematite photoanodes**

Hematite photoanodes having a nominal photoactive area of 100 cm$^2$ (see Figure S1A) were prepared by spray pyrolysis on fluorine-doped tin oxide (FTO) coated glass substrates, as detailed in the Experimental Procedures section. Current – potential (*I-E*) linear sweep voltammograms were measured in the dark and under solar simulated light (AM 1.5 G, 100 mW·cm$^{-2}$), and are presented in *Figure 6*. The long-term (1000 h) stability of similar photoanodes was demonstrated elsewhere.[21,31]

**Cathode**

A platinum coated titanium mesh electrode having a nominal area of 25 cm$^2$ was used as the cathode, see Figure S1B and additional details in the Experimental Procedures section. This cathode is suitable for demonstration purposes but is unlikely to be used in large-scale applications given the high cost of platinum. Such applications would require substituting the platinum with an Earth-abundant catalyst such as nickel-based catalysts for the alkaline hydrogen evolution reaction.[46]

**Photovoltaic cells**

Monocrystalline silicon photovoltaic (PV) mini-modules (89 mm × 55 mm) having a rated short-circuit current ($I_{SC}$) of 200 mA and open-circuit voltage ($V_{OC}$) of 5.04 V were purchased from IXYS (ref. SLMD481H08L IXOLAR™).[47] The PV mini-modules were characterized by current-voltage (*I-V*) scans and a stability test under solar simulated (AM 1.5 G, 100 mW·cm$^{-2}$) illumination (see Figures S2), as described in the Experimental Procedures section.

**Nickel hydroxide auxiliary electrodes**

Battery-grade nickel hydroxide pocket-plate electrodes (115 mm × 17 mm × 5 mm, weighting 19.2 g) were used as auxiliary electrodes (AEs). The electrodes' composition and phase were characterized by EDS and XRD, respectively, displaying an active material comprised of mainly β-Ni(OH)$_2$ (see Figures S3-S4). Their capacity and charge/discharge characteristics were evaluated by chrono-potentiometric cycles (see SI, Section S3.2).

## Decoupled photoelectrochemical system design criteria

### Operation principles

The separation of the oxygen and hydrogen cells is enabled by adding Ni(OH)$_2$-based AEs that mediate the ion (OH$^-$) exchange between the cathode and anode by undergoing a reversible redox reaction: Ni(OH)$_2$ + OH$^-$ ⇌ NiOOH + H$_2$O + e$^-$. During operation, the hydrogen evolution reaction (HER) takes place at the cathode, reducing water and releasing OH$^-$ ions (4 H$_2$O + 4 e$^-$ → 2 H$_2$ + 4 OH$^-$). In the conventional PEC cell configuration, the OH$^-$ ions diffuse through the electrolyte and separator to the photoanode surface, where they are oxidized by photogenerated holes to form oxygen (4 $h\nu$ → 4 e$^-$ + 4 h$^+$; 4 OH$^-$ + 4 h$^+$ → O$_2$ + 2 H$_2$O) through the oxygen evolution reaction (OER). In contrast, in our separate-cell configuration the OH$^-$ ions are consumed by the Ni(OH)$_2$ AE in the hydrogen cell (*Figure 1*B), oxidizing it to NiOOH (4 Ni(OH)$_2$ + 4 OH$^-$ → 4 NiOOH + 4 H$_2$O + 4 e$^-$). The electrons that are released in this charging reaction travel to the NiOOH AE in the oxygen cell through an external electrically-conducting wire, reducing it to Ni(OH)$_2$ (4 NiOOH + 4 H$_2$O + 4 e$^-$ → 4 Ni(OH)$_2$ + 4 OH$^-$). Finally, the OH$^-$ ions that are released by the discharging NiOOH AE in the oxygen cell reach the photoanode in the same cell, where they are oxidized by photogenerated holes (4 $h\nu$ → 4 e$^-$ + 4 h$^+$; 4 OH$^-$ + 4 h$^+$ → 2 H$_2$O + O$_2$). The ion (OH$^-$) exchange between the cells is therefore mediated by the OH$^-$ absorption and release through the AEs, thereby closing the electrical circuit within each cell wherein the two cells are connected electrically by electrically-conducting wires that conduct electrons from the photoanode to the cathode and from the NiOOH AE in the oxygen cell to the Ni(OH)$_2$ AE in the hydrogen cell.

The number of AEs in the hydrogen and oxygen cells is determined by the capacity of a single AE and the overall charge that should be transferred between the cells during one cycle. For a system operating at a photocurrent *I* for the duration *t*, a total charge of *Q* = *I* × *t* must be transferred between the oxygen and hydrogen cells by the AEs. If a single AE has a capacity of $Q_{AE}$ = $\alpha$ × *Q*, with 0 < $\alpha$ < 1 being a constant, the number of AEs that must be placed in each cell is *n* = 1/$\alpha$. Under our experimental conditions, which are limited by the photocurrent of the hematite photoanode, two AEs were required to perform daily cycles of 8 h (for detailed information see SI, Section S3.2).

After the AEs in the hydrogen cell had been charged and the AEs in the oxygen cell had been discharged, the AEs must be swapped between the cells to continue operation.

### Oxygen cell

The main structural aspect that distinguishes PEC cells from conventional electrolytic cells is the presence of an optically transparent window through which the photoelectrode is illuminated. In our separate-cell photoelectrochemical demonstration system the PEC-PV tandem device has a photoanode that produces oxygen whereas the hydrogen is produced in another cell that is physically separated from the oxygen cell. Therefore, only the oxygen cell must be transparent. Additionally, since only oxygen gas is produced in this cell, without hydrogen, there is no need to hermetically seal the cell or collect the product, as oxygen can simply be released to the atmosphere. This PEC cell design reduces the material and construction expenses for sealing and gas piping. Thus, the construction of large solar plants with many PEC cells is greatly simplified.

The main design criteria for our PEC oxygen cell are:

(i) Simplicity of construction, operation and maintenance;
(ii) Abundant, cheap and durable construction materials and components;
(iii) Light penetration though the cell to the photoanode and to the PV cell behind it;
(iv) The cell must host at least one photoanode and at least one nickel oxyhydroxide (NiOOH) AE;
(v) The photoanode is connected electrically to a PV mini-module that provides the necessary bias for unassisted solar water splitting;
(vi) The AE in the oxygen cell is connected electrically with the AE in the hydrogen cell.

***Figure 3*** presents the assembled oxygen cell and its components. First, to reduce construction costs and complexity, the oxygen cell geometry must be as simple as possible, avoiding the need for screws, O-rings and other construction parts. Considering the geometry imposed by the flat photoanode, the oxygen cell was constructed as a rectangular container. Since the oxygen evolved can be released to the atmosphere, this cell does not need hermetic sealing, as explained above. A removable lid was placed over the top of the container to minimize electrolyte evaporation, $CO_2$ poisoning, and contamination from the environment (*e.g.* sand, dust, *etc.*). The lid was fitted with a water inlet tube to compensate for any water loss and control the electrolyte level (see Figure S9, Section S4.1.3 in the SI). The removable lid allows for easy insertion of the AEs in the oxygen cell, and thus reduces the complexity of the AE swapping operation. In a future commercial installation, the electrolyte may be circulated with an inlet from the bottom and an outlet from the top, ensuring complete filling of the electrode space, as well as actively pushing the oxygen bubbles upward and out of the vessel. Furthermore, in a well-designed system, the solar heating in the oxygen cell can be used to drive a thermal syphon mode, avoiding the need for electrical pumping and increasing the overall system efficiency.

➔Figure 3

Next, the optical, physical and chemical properties of the oxygen cell construction material must be carefully considered not only for this demonstration but also considering the scale-up potential of the system and its cost. Specifically, the cell construction material must have high optical transmittance in the relevant wavelengths (350 – 1200 nm, covering both the hematite photoanode and the Si PV cells behind it) while maintaining a high resistance to chemical attack by the alkaline solution at moderate temperatures (up to 50°C)[48] under direct sunlight exposure. Poly(methyl methacrylate) (PMMA), also known as Plexiglass, was the material of choice for the cell construction. PMMA is a low-cost material, which is optically transparent at the relevant wavelengths, typically with a visible light transmission of up to 92% for 3 mm thickness.[49] Moreover, PMMA is chemically and mechanically stable in aqueous alkaline solutions, as well as under prolonged sunlight exposure.[38,50] For all current collectors, 316 stainless-steel was used due to its excellent stability and corrosion resistance[51] (see Figure S10, Section S4.1.4 in the SI).

Finally, the component placement within the cell and the cell dimensions were optimized, as discussed in detail in the SI (Section S4.1, Figures S6, S7 and S8). Accordingly, the oxygen cell was designed to host: (i) a single hematite photoanode (100 cm$^2$) or two photoanodes stacked back-to-back; (ii) Si PV mini-modules of a total area not larger than the photoanode; and (iii) two NiOOH auxiliary electrodes for performing daily cycles of 8 h. The AEs were placed besides the photoanodes, perpendicular to the PEC-PV panel, as in a side-by-side cell configuration (see Figure S7 – Design D, Section 4.1 in the SI).

**Hydrogen cell**

The hydrogen cell must be:

(i) compact;
(ii) hermetically sealed;
(iii) made of durable construction material and components.

Unlike the PEC oxygen cell, the hydrogen cell does not have to be transparent, so it may resemble conventional alkaline electrolyzes. However, there are two notable differences between the hydrogen cell in our system and an alkaline electrolyzer. Firstly, there is no need for a membrane or diaphragm in our hydrogen cell because the oxygen evolution occurs elsewhere, in another cell (the oxygen cell); and secondly, the cell must be periodically opened and resealed to swap the AEs between the two cells. While the lack of separator simplifies the cell design and construction, the AE swapping complicates its operation. Nevertheless, the swapping frequency between the two cells can be reduced and the operation simplified by designing the hydrogen cell to accommodate a reserve stock of extra AEs. This is explained in detail in the following section.

Although the hydrogen cell can be opaque, in this work a transparent glass cell was constructed for demonstration purposes, so that the hydrogen gas bubbles and cell components would be visible (Figures S11 and S12, Section S5 in the SI). The cell allows for simple continuous operation at a benchtop scale application in ambient pressure. For long-term operation, the cell must be robust, and its construction material should be chemically and mechanically stable in alkaline conditions. Figure 4 presents the assembled hydrogen cell and its components.

➔Figure 4

**Auxiliary electrode swapping frequency**

While the oxygen cell is a simple container allowing for easy AE management, the hydrogen cell is a hermetically sealed container. Handling the AEs in the hydrogen cell is therefore more complicated than their counterpart AEs in the oxygen cell. It is therefore favorable to reduce the AE swapping frequency. As mentioned previously, the swapping frequency between the cells can be reduced by designing the hydrogen cell to accommodate a reserve stock of extra AEs that are operated sequentially such that at any given time some AEs are operating whereas the rest of them are in idle state. Unlike the oxygen cell, where the number of AEs that can be placed in front of the photoanode is limited due to light blocking, multiple AEs can be placed in the hydrogen cell while still maintaining a compact design. Only when all the AEs in the reserve stock are fully charged, the entire stack of AEs is swapped with their counterpart AEs in the oxygen cell.

*Figure 5* illustrates the proposed AE swapping scheme, where $Y$ discharged AEs are placed inside the hydrogen cell, out of which only $X$ ($X < Y$) AEs are connected to the oxygen cell. Similarly, a reserve stock of $Y$ charged AEs is placed at the solar site, out of which only $X$ AEs are placed in the oxygen cell and connected to the hydrogen cell, as illustrated in *Figure 5*A.

→Figure 5

After one cycle of operation, the *X* (now discharged) AEs are removed from the oxygen cell and stored, and *X* other charged AEs from the reserve stock are introduced to the oxygen cell in their place. This is termed *intracell swapping,* where only the AEs in the oxygen cell are collected and swapped with AEs from the same stock (**Figure 5**B). For the next cycle, *X* (now charged) AEs in the hydrogen cell are disconnected, and a different set of *X* "fresh" (i.e., discharged) AEs are connected in their place, without opening the hydrogen cell (**Figure 5**C). Only after all *Y* electrodes in the reserve stock have been exhausted, after $n = Y/X$ *intracell swaps*, all the discharged AEs from the solar field are collected and swapped with all their charged counterparts from the hydrogen cell. This is termed *intercell swapping*, and its frequency can be controlled by modifying the ratio of *Y/X* (**Figure 5**D). This operation scheme is designed to reduce the *intercell swapping* frequency. For the purpose of this demonstration, a daily *intracell swapping* frequency was chosen where a single day is considered as 8 h (the average number of sunlight hours in a day) and the swapping is performed overnight, with *X* = 2 and *Y* = 4, such that the *intercell swapping* occurs every other day.

## Results and discussion

### Tandem PEC-PV device optimization

In a dual photoabsorber PEC–PV tandem configuration, the photoanode and the PV cell use photons of different spectral ranges of the solar spectrum to enable optimal sunlight harvesting. The PV is electrically connected in series between the photoanode and the counter electrode. Due to the series connection, the current is limited by the cell with the lowest current, which is typically the PEC cell. It is therefore desirable for the individual photocurrent responses of the PEC and PV cells to match each other in order to minimize the electrical coupling loss that arises from series connection of the two cells.

Figure 6A shows *I-E* linear sweep voltammograms (LSV) of two hematite photoanodes, measured in front and back side illumination as well as in the dark. Both samples presented similar photocurrent for all applied potentials in dark and under front side illumination, with an onset potential of 1 $V_{RHE}$, and achieving a photocurrent of 34 mA at 1.4 $V_{RHE}$ and 64 mA at 1.8 $V_{RHE}$. After characterization of both samples separately, they were connected back-to-back such that sample *I* was front illuminated, and sample *II* was back illuminated (placed behind sample *I*), with their photoactive hematite layers facing out toward the electrolyte solution. The *I-E* voltammogram of this photoanode stack is shown by the black solid curve, yielding a superposition of the front and back illuminated curves. This resulted in a photocurrent of 43 mA at 1.4 $V_{RHE}$, which corresponds to a 26% current increase compared to a single photoanode illuminated from the front.

→Figure 6

*I-V* characteristic curves of the Si PV mini-module measured under simulated solar illumination (AM 1.5 G, 100 mW·cm$^{-2}$) are presented in Figure 6B. The PV module displays a short-circuit current ($I_{SC}$) of 164.5 mA and an open-circuit voltage ($V_{OC}$) of 5.4 V. However, after placing the PV module against the back window of the oxygen cell (filled with 1 M NaOH solution), the current decreased by 20%. When the two back-to-back photoanode stack was placed in the cell, the $I_{SC}$ was 59 mA (a current decrease of 64% relative to the directly illuminated module). The crossover point between the *I-E* voltammogram of the hematite photoanode stack and the *I-V* curve of the PV module placed behind the oxygen cell (with the hematite photoanode stack in it) yields a predicted operation point with a photocurrent of 58 mA at a voltage of 1.7 V. It is noted that this crossover point does not fully account for all the voltage losses in the entire system. Specifically, it does not account for Ohmic and polarization losses in the electrolyte, the counter electrode (*i.e.*, the cathode) and the auxiliary electrodes.[52] However, these losses are smaller than the anode polarization loss and therefore the crossover point provides a fair approximation of the operation point of the entire system.

The performance of the PV module and the PEC-PV tandem system proved to be stable over 8 h, with no current decrease (Figure S2). It is noted that similar hematite photoanodes were tested for 1000 h displaying stable photocurrent with no sign of degradation, as reported elsewhere.[21,31] During water splitting, changes in the AEs state-of-charge (SOC) and bubble accumulation may shift the system voltage, and thus shift the operating point. Hence, if the operating point of the system is too close to the PV module's maximum power point (MPP), a shift in the system voltage could cause the current to decrease, resulting in operation instability. The Si PV module was selected accordingly, for optimal operation in tandem with the hematite photoanode stack, such that the operating voltage is well below the PV module's MPP, and below the onset potential of the dark current of the photoanode stack (1.8 $V_{RHE}$; see Figure 6A).

**Auxiliary electrodes optimization**

The standard Ni(OH)$_2$/NiOOH redox potential is 1.42 $V_{RHE}$,[53–55] higher than the reversible OER potential (1.23 $V_{RHE}$). However, the actual reversible reaction potential depends on the electrode's SOC, composition, and the pH and composition of the solution,[53,55] and may therefore be lower than 1.42 $V_{RHE}$. Since the OER has a high overpotential (300 – 500 mV at 10 mA·cm$^{-2}$),[56] the Ni(OH)$_2$ electrode's SOC range and composition can be tailored to enable charging without concurrent oxygen evolution. Similarly, the electrode can be readily discharged without hydrogen evolution if full discharge is avoided, allowing for complete product separation without co-generation of hydrogen and oxygen in the same cell.[40] To select the appropriate SOC operation window, several parameters must be considered:

(i) the overall charge, which is calculated as the product of the predicted operating current and the required operation time ($Q = I \times t$);
(ii) the extent of the parasitic OER during charging, which increases with SOC and electrode potential;
(iii) the extent of the HER during discharging, which occurs if the electrode is reduced beyond full discharge;
(iv) the stability of the AEs throughout the operation (full charge/discharge should be avoided).

The overall charge that must be transferred in each cycle is given by the product of the operation current (predicted to be 57.6 mA, see Figure 6B) and the required operating time per cycle. Here, a single day operation was taken as a benchmark, equivalent to 8 h per cycle (commensurate with the average number of sunlight hours in a day). Thus, the AEs must transfer a charge of 460 mAh, which is 38% of their full capacity (1200 mAh, see SI, Section S3.2 and Figure S5A).

To choose the optimal SOC range limits, two effects were considered. On the one hand, raising the SOC upper limit increases the extent of the parasitic OER at the charging AE in the hydrogen cell. On the other hand, decreasing the SOC lower limit increases the possibility that fluctuations in operation will lead to full discharge, followed by hydrogen evolution at the AE in the oxygen cell. Yet a high SOC limit means a high AE capacity, and is desired for longer operation and less frequent swapping. Therefore, the operation conditions should be optimized with respect to these two effects, for a maximized SOC range. To balance these effects, the extent of the parasitic OER was measured at different SOC limits (see SI, Section S3.2, Figure S5B). For a single AE, significant OER was observed at a SOC of 40%, corresponding to 480 mAh. To avoid oxygen generation in the hydrogen cell, two AEs were used in each cycle, such that each AE transfers a charge of only 240 mAh (20% of its full capacity). The optimal SOC range in this configuration was selected between 8% and 28% of the full capacity of the electrode, corresponding to a range of 100 – 340 mAh. The extent of the parasitic OER in this range was estimated to be 4% of the total passed charge, equivalent to an oxygen volume fraction of 1.96%, far below the combustion limit of oxygen in hydrogen (6 vol.%) (see SI, section S3.2).

**Decoupled photoelectrochemical system characterization**

The separate-cell PEC system was assembled with the optimized PEC-PV tandem device, the AEs and a cathode. The oxygen cell was assembled as presented in ***Figure 3***, with: (i) a photoanode stack comprised of two hematite photoanodes (100 $cm^2$) connected back-to-back; (ii) one Si PV module (50 $cm^2$) placed outside the cell behind the photoanode stack, and; (iii) two AEs, each charged to 340 mAh, placed inside the cell perpendicular to the photoanode stack. The arrangement of the AEs within the cell was optimized to maximize the solar to hydrogen (STH) conversion efficiency, as described in detail in the SI (Section S4.1). Two additional (charged) AEs were kept in 1 M NaOH electrolyte as a reserve stock for the oxygen cell. The hydrogen cell was assembled as presented in Figure 4, with: (i) a single cathode made of platinum-coated titanium mesh (25 $cm^2$), and; (ii) four AEs, each charged to 100 mAh, out of which only two were connected to the electric circuit while the other two remained in the reserve stock. Both oxygen and hydrogen cells were filled with 1 M NaOH aqueous solution and kept at ambient conditions without temperature control.

Prior to operation, the system's individual components were characterized separately. First, the HER and OER potentials of the cathode and anode were examined by LSV of the cathode and hematite photoanode, respectively (Figure 7A). Next, the open circuit potentials (OCP) of the AEs in each cell were measured, yielding 1.36 $V_{RHE}$ for the discharged AEs in the hydrogen cell, and 1.32 $V_{RHE}$ for the charged AEs in the oxygen cell.

Then, each cell was characterized separately as a stand-alone unit (Figure 7B). On its own, the hydrogen cell operates similarly to charging a Ni-$H_2$ battery, wherein the Ni(OH)$_2$ anode is charged to NiOOH while hydrogen is evolved at the cathode. The current onset for the hydrogen cell was observed at 1.32 V, in accordance with the data presented in Figure 7A for discharged AEs OCP (1.32 $V_{RHE}$) and the HER current onset at the cathode (~0 $V_{RHE}$). The oxygen cell was examined in the dark and under solar-simulated illumination (1 Sun, 100 mW·$cm^{-2}$). Here, without illumination the OER proceeds at the photoanode stack at $E > 1.6$ $V_{RHE}$ while the AEs are discharged at $E < 1.36$ $V_{RHE}$ (Figure 7A). Accordingly, the current onset for the entire cell was 0.25 V. However, under illumination, the OER onset potential at the photoanode stack is 0.8 $V_{RHE}$, lower than the AEs OCP. Therefore, under illumination the oxygen cell alone operates spontaneously as a discharging "Ni-$O_2$" battery, until the potential of the discharging AEs drops below the photoanode's OER potential.

Finally, the two cells were connected to each other by connecting the two AEs in the oxygen cell with two AEs in the hydrogen cell via a copper wire, with the photoanode and cathode connected to a potentiostat as the working and counter electrodes, respectively. The *I-V* characteristic curve for the separate-cell system is presented in Figure 7C, overlaid on the PV module's *I-V* curve. Under illumination, the predicted operating point of system was *I* = 56.4 mA at *V* = 1.86 V, close to the predicted operating point based on the photoanode stack's performance shown in Figure 6B.

➔Figure 7

**Decoupled PEC water splitting demonstration with simulated sunlight (indoors)**

To demonstrate overall solar water splitting under controlled lab conditions, the two AEs in the oxygen cell were connected to two of the four AEs in the hydrogen cell, and the photoanode stack was illuminated with a solar simulator (calibrated to 1 Sun) without any external bias, while measuring the short-circuit current between the photoanode stack in the oxygen cell and the cathode in the hydrogen cell. The first cycle was stopped when the transferred charge reached a pre-calculated value of 456 mAh (57 mA × 8 h) and lasted for 8.3 h at an average current of 55 mA.

It is important to note that unlike Ni-$H_2$ batteries, where the Ni(OH)$_2$ anode is charged and discharged in separate steps, the AEs in the separate-cell system are charged and discharged simultaneously. In this scheme, the amount of charge transferred to the charging AEs in the hydrogen cell is limited by the amount of charge transferred from the discharging AEs in the oxygen cell. Incomplete charging of the AEs in the oxygen cell due to parasitic OER will therefore lead to charge accumulation that may eventually result in operation instability. However, this can be amended by extending the operation of the hydrogen cell between cycles and recharging the AEs to their initial state, as shown in our previous work.[40] Accordingly, after the first water splitting cycle was completed the hydrogen cell was disconnected from the oxygen cell and operated separately (using an external power source) at a constant voltage of 1.45 V for an additional 45 mAh for drift-free water splitting cycles.

For the second cycle, the two (now discharged) AEs were taken out of the oxygen cell and swapped with two (charged) AEs from the reserve stock. The two (now charged) AEs in the hydrogen cell were disconnected, and the other two (discharged) AEs in that cell were connected instead. The second cycle was performed in the same way and lasted 8.3 h at an average current of 54 mA. Again, the hydrogen cell was disconnected after this cycle and operated at 1.45 V. After this cycle, the four (now discharged) AEs from the oxygen cell and the reserve stock were swapped with the four (now charged) AEs from the hydrogen cell, and the operation continued. Figure 8A shows stable short-circuit current traces measured during ten water splitting cycles carried out using the separate-cell PEC demonstration system, with an average cycle duration of 8.2 h, at an average short-circuit current of 55.2 mA. The specific charge that was transferred to the AEs in each cycle was 55.2 mA × 8.2 h / 2 electrodes / 19.2 g/electrode = 11.7 mAh·$g^{-1}$. Similar electrodes were examined elsewhere,[40] demonstrating high Faradaic efficiency (~100%) during charging and discharging of 11 mAh·$g^-$ [1] (see Figure S19 in Ref. [40]). Gas chromatography measurements displayed pure hydrogen generation with no concurrent oxygen generation in the hydrogen cell during charging to a specific charge of 62 mAh·g$^{-1}$ (see Figure S21 in Ref. [40]).

➔Figure 8

Figure 8B shows the system voltage ($V_{sys}$), as well as the potentials of the charging and discharging AEs ($E_{AE,\text{hydrogen cell}} = E_{AE,H_2}$ and $E_{AE,\text{oxygen cell}} = E_{AE,O_2}$, respectively) throughout a single water splitting cycle. For 8 h operation period the average system voltage was 1.83 V, corresponding to a voltage efficiency of $\eta_V = 1.23\text{ V}/1.83\text{ V} = 67\%$. The average current and voltage for each cycle is in good agreement with the predicted operating point, identified as (×) in Figure 7C.

It is noteworthy that this separate-cell water splitting system can be viewed as two cells connected in series, the characteristics of which are presented in Figure 7B. Therefore, the total system voltage can be written as the sum of the two cell voltages:

$$\begin{aligned} V_{\text{sys}} &= V_{H_2\text{ cell}} + V_{O_2\text{ cell}} = \\ &= \left[(E_{AE,H_2} - E_{\text{cathode}}) + \sum IR_{H_2}\right] + \left[(E_{PA} - E_{AE,O_2}) + \sum IR_{O_2}\right] + IR_c \\ &= (E_{PA} - E_{\text{cathode}}) + (E_{AE,H_2} - E_{AE,O_2}) + \sum IR_{H_2} + \sum IR_{O_2} + IR_c \\ &= \Delta E_{PA-\text{cathode}} + \Delta E_{AE} + \sum IR \end{aligned}$$

where $V_{H_2\text{ cell}}$ and $V_{O_2\text{ cell}}$ are the individual cell voltages of the hydrogen and oxygen cells, respectively, $\sum IR_{H_2}$ and $\sum IR_{O_2}$ are the Ohmic losses of each cell, $IR_c$ is the Ohmic loss associated with the wire connecting the two cells, $E_{PA}$ and $E_{\text{cathode}}$ are the photoanode and cathode potentials, respectively, $\Delta E_{PA-\text{cathode}}$ is the potential difference between the photoanode and cathode, and $\Delta E_{AE}$ is the potential difference between the charging and discharging AEs. Neglecting the Ohmic losses of the additional circuit elements and electrolyte mass, the voltage efficiency of the separate-cell system compared to an equivalent single-cell system is directly influenced by the potential difference between the charging and discharging AEs, $\Delta E_{AE}$. As shown in Figure 8B, the average $\Delta E_{AE}$ during this water splitting test was 100 mV. The voltage of an equivalent single-cell system, employing the same photoanode and cathode, can be calculated as 1.83 V − 0.1 V = 1.73 V. Thus, the voltage efficiency of this separate-cell system is 95% of the voltage efficiency of an equivalent single-cell system.

The STH efficiency ($\eta_{STH}$), determined from Equation S5 (Section S2.1.2 in SI),[57] of the PEC-PV tandem system for all ten cycles was 0.68%, taking $\eta_F$ = 100% and $I_{SC}$ = 55.2 mA. The highest current achievable for an equivalent single cell system is 57.6 mA, obtained at the crossing point of the photoanode stack and PV module's *I-V* curves (Figure 6B), yielding $\eta_{STH}$ = 0.71%. Hence, similarly to the voltage efficiency comparison, cell separation reduces the STH efficiency by only 5%, as compared to an equivalent single-cell system.

**Decoupled PEC water splitting demonstration with natural sunlight (outdoors)**

The performance of the separate-cell PEC water splitting system was evaluated at outdoor environmental conditions under natural sunlight. The setup was assembled as in the indoor demonstration, without any additional bias and without the solar simulator. The oxygen cell was placed in an open space exposed to the sunlight (without shade) and tilted to an angle of 41° relative to the horizon, facing south (***Figure 9***A). This experiment was carried out in Haifa,

Israel on October 4[th] 2018 between 10:20 AM and 6:20 PM, while the weather was partly cloudy. *Figure9*B shows the current measured throughout this 8 h experiment. The average current was 31.6 mA. However, the generated current was > 50 mA up until 2:00 PM and only after this hour it started to decrease due to the meteorological conditions, such as lower solar radiation and cloudiness.

➔Figure 9

**Simultaneous hydrogen and electricity generation**

The *I-V* curves of the PV module and the separate-cell PEC system were designed to intersect just before the PV module's MPP for stability reasons, as explained above, and before the onset of the dark current, in which the system behaves like an electrolytic cell. The PV module selected to comply with these specifications had an area of 50 cm$^2$, which equals half of the photoanode stack's area. It is possible to utilize the entire illuminated area of 100 cm$^2$ by connecting two of such PV modules, either in parallel or in series.

Figure *10*A presents the *I-V* characteristic curves of the possible PEC-PV configurations. When two PV modules were connected in series, the *I-V* curve of the PV modules crosses that of the PEC system at the same operating point as a single PV module. Therefore, while the PV modules conformally map the PEC illuminated area, this configuration is wasteful. When the two PV modules were connected in parallel, mapping the PEC system's illuminated area, the operating point is obtained at 2.5 V, above the onset potential of the dark current; above this potential (> 1.8 V), hydrogen production is not purely photoelectrochemical but electrolysis also takes place. It is possible to utilize the entire illuminated area with two PV modules connected in parallel by adding a secondary load in parallel to the PEC system in order to produce electrical power in addition to hydrogen, as proposed elsewhere.[15] To demonstrate such a scenario, a 25 Ω auxiliary load was connected in parallel to the separate-cell PEC device. The predicted operating point of this new arrangement is at a *V* = 1.7 V with the highest current of *I* = 115.4 mA, which is divided between the load (68 mA) and the PEC system (47.4 mA).

➔Figure 10

Figure *10*B shows the result of a separate-cell PEC water splitting cycle with a 25 Ω load connected in parallel to the PEC system. The average system voltage throughout the cycle was 1.74 V, and the system produced hydrogen at an average current of 45.8 mA, *i.e.*, $\eta_{STH}$ = 0.56%, as well as a net electrical power of 68.5 mA × 1.74 V = 119.2 mW, corresponding to a power conversion efficiency of 1.2%. The overall efficiency is therefore 1.76%, 2.5 times higher than that of the separate-cell PEC system with a single PV module without an external load.

**Conclusions**

This work demonstrates decoupled solar water splitting in separate oxygen and hydrogen cells. The separation of hydrogen and oxygen production into two cells connected electrically, with nickel hydroxide / oxyhydroxide AEs serving as ion exchange mediators, offers important advantages over conventional PEC cell architectures, as reported elsewhere.[40] Key advantages include: (i) avoiding the need of membrane separators; (ii) eliminating the need for installing a piping manifold and sealing the solar panels, since only the oxygen cell has to be placed at the solar site and the oxygen evolved at the photoanode can be released to the atmosphere; (iii) centralized hydrogen production, and; (iv) enabling easy upscale and safe operation.

The PEC-PV tandem configuration was optimized to maximize the system efficiency, as well as the AEs cycling process and cells arrangement for reducing the AEs swapping frequency and the system cost. The oxygen cell contained a 100 cm$^2$ hematite photoanode back-to-back stack in tandem with a Si PV module, and two NiOOH AEs. The hydrogen cell is a regular electrolytic cell comprising a single platinized cathode and four Ni(OH)$_2$ AEs, out of which only two were connected to the electric circuit for each operation. The system performance was assessed carrying out ten stable water splitting cycles (each cycle duration was 8.3 h) at an average current of 55.2 mA and a solar to hydrogen efficiency of 0.68%. The viability of the separate-cell PEC system was demonstrated under natural sunlight in outdoor conditions reaching similar performances to lab-indoor tests. Finally, an efficiency improvement of 2.5 times was achieved by connecting a 25 Ω load in parallel to the PEC system for simultaneous hydrogen and electricity generation, yielding an overall efficiency of 1.7%.

To the authors' knowledge, this is the first reported stable separate-cell PEC water-splitting system at this scale. A previous study reported decoupled PEC water splitting using a soluble redox mediator (H$_3$PMo$_{12}$O$_{40}$/H$_5$PMo$_{12}$O$_{40}$) in an acidic electrolyte solution (pH = 0.1).[58] In this system, the PEC cell had two compartments separated by a Nafion membrane, similar to the conventional PEC cell architecture. Under chopped-light solar-simulated illumination (10 pulses of 30 s), the photoanode (WO$_3$, 1 cm$^2$) oxidized water (generating oxygen) while the soluble redox mediator was reduced at the cathode from H$_3$PMo$_{12}$O$_{40}$ to H$_5$PMo$_{12}$O$_{40}$ through a proton-coupled electron transfer reaction, at a photocurrent density of 1.2 mA·cm$^{-2}$. Then, the reduced catholyte was transferred to an electrolytic cell in which it was oxidized, under a bias voltage of *ca.* 1.5 V (obtained using an external power source), releasing hydrogen and regenerating the redox mediator.

This approach has the advantage of using a soluble redox mediator that can be easily transferred between the two cells (oxygen and hydrogen cells),[58] as opposed to our proposed approach that employs a solid-state redox mediator, the Ni(OH)$_2$/NiOOH AEs, that must be swapped mechanically. This is the biggest technical challenge in our approach and would most likely require an automated system to collect the AEs from the PEC oxygen cells. Nevertheless, the voltage efficiency of the decoupled (separate-cell) system with respect to the equivalent single cell system in which the oxygen and hydrogen evolution reactions remained coupled is only 79% using the H$_3$PMo$_{12}$O$_{40}$/H$_5$PMo$_{12}$O$_{40}$ redox mediator, as reported in another study of the same system,[43] whereas in our case it is 95%, as reported in this work and in our previous study.[40] Likewise, the Faradaic efficiency of the redox mediator charging reaction is lower for the H$_3$PMo$_{12}$O$_{40}$/H$_5$PMo$_{12}$O$_{40}$ couple (84% ± 6%)[58] compared to the Ni(OH)$_2$/NiOOH AEs (96%). Therefore, our approach offers a considerably higher efficiency. Another advantage of our approach is the membrane-less operation, unlike the alternative approach based on a soluble redox mediator, which requires membranes in both the oxygen and hydrogen cells.[43,58] Finally, all of our cell components are suitable for stable operation in an aqueous alkaline electrolyte that is transparent in the UV-visible range, enabling a PEC-PV tandem design that provides all the power for the overall water splitting reaction, as demonstrated in this work. By contrast, the electrolyte in the alternative approach is colored by the H$_3$PMo$_{12}$O$_{40}$/H$_5$PMo$_{12}$O$_{40}$ couple,[43,58] which prevents placing a PV cell behind the PEC cell. As a result, an external power source must be used to provide additional bias voltage to complete the overall water splitting reaction.[58] Thus, each approach, using solid-state vs. soluble redox mediators, has pros and cons that must be carefully balanced when selecting a decoupling strategy for PEC water splitting.

Besides the specific challenges related to the decoupling strategy, another important challenge lies in improving the performance of stable photoanodes, such as the hematite photoanodes used in this study, for successful integration with silicon (or other) PV cells for efficient PEC-PV tandem cells that could eventually compete with PV-electrolysis systems.[15] For pushing the PEC water splitting closer to marketable application, further efforts should focus on improving the performance of stable photoelectrodes.

**Experimental procedures**

**Hematite photoanodes fabrication**

Hematite thin films of *ca.* 20 nm were deposited on 2.2 mm thick, 7 Ω·square$^{-1}$ conducting fluorine-doped tin oxide (FTO) coated glass substrates (Solaronix TCO 22-7, Switzerland) by spray pyrolysis using the UPORTO assembled setup.[21] First, the FTO substrates (110 mm × 100 mm) were cleaned by repeated ultrasonic treatments in soapy water, acetone, ethanol, 1 M KOH solution and deionized water (15 min each). Before deposition, a conductive silver contact (Ferro Paste GmbH GSSP SP 1963, USA) was manually printed on the substrate's top edge and sintered at 500 °C,[31] and the substrate active area (100 mm × 100 mm) was then pre-treated with *ca*. 30 mL of a diluted tetraethyl orthosilicate (TEOS) solution (99.9 %, Aldrich; 10% volume in ethanol) at 450 °C using a hand-spray glass atomizer.[21] A gold layer of 100 nm was deposited on the silver contact using an electron-beam evaporator (Airco Temescal FC 1800) to protect the silver contact from oxidation. The hematite deposition conditions on TEOS treated FTO substrates were optimized as reported elsewhere.[31] Briefly, the substrates were placed over a heating plate at 450 °C and the spray nozzle placed 30 cm over the substrates was fed with an ethanolic solution of iron and mixed with compressed air. An automatic syringe pump (Cronus Sigma 2000 Dual Syringe Pump, UK) was used to deliver 1 mL of 10 mM of Fe(acac)$_3$ (99.9+%, Aldrich) at 12 mL·min$^{-1}$. The time gap between sprays was 45 s and the total solution volume applied was 70 mL. The hematite photoelectrodes were air-annealed for 30 min at 500 °C, before being cooled down to the room temperature. Figure S1A shows an image of the hematite photoanode prepared with an active area of 100 cm$^2$.

**Hematite photoanodes photoelectrochemical characterization**

Prior to the photoanodes' characterization, the solar simulator lamp's stability was verified by placing a Si PV module having a rated $I_{SC}$ of 50 mA and a $V_{OC}$ of 2.52 V (ref. SLMD121H04L IXOLAR$^{TM}$ High Efficiency SolarMD monocrystalline solar cell 44 mm × 14 mm, IXYS)[59] under illumination and measuring the short-circuit current for 8 h. Figure S2 shows the short-circuit current traces of the Si PV module and a PEC-PV tandem system, displaying stable operation over the test duration (8 h). The PEC-PV tandem measurement was conducted in a Pyrex cell using two 100 cm$^2$ hematite photoanodes which were coupled with three PV modules connected in parallel.

The photoelectrochemical performance of the hematite photoanodes (Figure S1A) was evaluated using an Ivium Potentiostat / Galvanostat workstation (Vertex Potentiostat / Galvanostat, Ivium Technologies). All the electrochemical characterization experiments were carried out in a standard three-electrode configuration: each hematite photoanode was connected to the potentiostat as the working electrode, with a titanium platinized mesh (*ca.* 25 cm$^2$) counter electrode and an Hg/HgO/1 M NaOH reference electrode (RE6-AP reference electrode for alkaline solution, ALS Co.). All components were immersed in 1 M NaOH aqueous electrolyte solution (pH ≈ 14). Special care was taken to avoid dipping the photoanode's electric contact in the electrolyte solution. The linear sweep voltammetry (LSV) curves presented in Figure 6A were obtained by applying an external bias to the photoanode and measuring the generated current in the dark and under solar simulated illumination using a class AAA solar simulator (Sun 3000, ABET Technologies) with a 150 W Xe lamp and an AM 1.5 G filter; the intensity was calibrated to 1-Sun (100 mW·cm$^{-2}$) using a reference PV cell (AK series, Konica Minolta Optics, INC). The potential scan rate was 20 mV·s$^{-1}$.

**Platinized titanium mesh cathode electrochemical characterization**

A titanium mesh electrode with double-sited platinum coating having a nominal surface area of 25 cm$^2$ (Baoji Longsheng Non-Ferrous Metal Co., Ltd) was used as the cathode (Figure S1B). The LSV curve for this cathode presented in Figure 7A was obtained in a three-electrode configuration with a nickel foil counter electrode (25 cm$^2$) and an Hg/HgO/1 M NaOH reference electrode in 1 M NaOH aqueous solution.

**Photovoltaic cells photoelectrical characterization**

Monocrystalline silicon photovoltaic (PV) modules were purchased and characterized under 1-Sun illumination. PV modules having a rated short circuit current ($I_{SC}$) of 200 mA and an open circuit potential ($V_{OC}$) of 5.04 V (ref. SLMD481H08L IXOLAR™ High Efficiency SolarMD monocrystalline solar cell 89 mm × 55 mm, IXYS),[60] were used for the experiments described in the article. The *I-V* measurements of the PV module, presented in Figure 6B, were obtained at a scan rate of 20 mV·s$^{-1}$.

**Nickel hydroxide AEs phase and composition analysis**

A battery-grade nickel hydroxide auxiliary electrode (Zhuhai Seawill Technology Co., Ltd) was disassembled and analyzed by X-ray diffraction (XRD) and Energy-dispersive X-ray spectroscopy (EDS). The XRD analysis (Figure S3) shows that the electrode comprises mostly of graphite and nickel hydroxide in the β-phase. The elemental composition was verified by EDS, summarized in Figure S4.

**Nickel hydroxide AEs electrochemical characterization**

Battery grade nickel hydroxide pocket-plate electrodes (115 mm × 17 mm × 5 mm, Figure S1C), weighting 19.2 g, were obtained from Ni-Fe batteries (Zhuhai Seawill Technology Co., Ltd), and used as auxiliary electrodes (AEs). All the characterization experiments were carried out in a three-electrode configuration with a nickel foil counter electrode (25 cm$^2$) and an Hg/HgO/1 M NaOH reference electrode immersed in a 1 M NaOH electrolyte solution. Chrono-potentiometric charge-discharge tests for capacity estimation were carried out at 100 mA. All other charge-discharge tests were carried out at a charging current of 30 mA and discharging current of 5 mA. Further details are described in the Supplemental information (Section S3.2).

**Acknowledgment**

This project has received funding from the European Research Council (ERC) under the European Union's Seventh Framework programme (FP/2007-2013)/ERC (grant agreement number 617516), the European Research Council under the European Union's Horizon 2020 research and innovation programme (grant agreement number 825117), and the PAT Center of Research Excellence supported by the Israel Science Foundation (grant no. 1867/17). P. Dias acknowledges the Portuguese Foundation for Science and Technology (FCT), I.P. for her postdoc scholarship (SFRH/BPD/120970/2016), and projects SunStorage (POCI-01-0145-FEDER-016387) and LEPABE-2-ECO-INNOVATION (NORTE-01-0145-FEDER-000005), funded by ERDF, through COMPETE2020 - OPCI, by FCT.

**Author Contributions**

A. L. designed the research and analyzed the data. R. W. performed the main experiments and analyzed the data. P. D. prepared and characterized the hematite photoanodes and assisted in the design of the oxygen cell. H. D., A. M., and G. E. S. assisted in experimental and system design. M. H. and O. N. performed characterization tests of the auxiliary electrodes. A. L., P. D., A. M., A. R., and G. S. G. wrote the manuscript. A. R. and G. S. G. supervised and guided the project.

**Declaration of Interests**

A.L., H.D., G.E.S., G.S.G. and A.R. hold the following patent in relation to decoupled water splitting: United States Patent No. 10,487,408 "Methods and system for hydrogen production by water electrolysis". H.D., G.S.G. and A.R. are co-founders of H2Pro Ltd. and are members of its board of directors. H.D. is the CTO of H2Pro Ltd. G.S.G. and A.R. are advisors of H2Pro Ltd.

**Main figures titles and legends:**

*Figure 1: Photoelectrochemical water splitting cell architectures.* (A) Conventional single-cell configuration of a PEC cell comprising a photoanode-PV tandem stack and cathode, separated by a membrane or diaphragm. (B) Separate-cell configuration for decoupled PEC water splitting with an oxygen-producing tandem PEC-PV cell and a hydrogen producing electrolytic cell connected to each other electrically.

*Figure 2: Conceptual illustration of a solar hydrogen refueling station with distributed PEC solar cells producing oxygen and a centralized hydrogen generator.* Reproduced from Ref. [40], with permission (Copyright © 2017, Springer Nature).

*Figure 3: Photoelectrochemical oxygen generation cell.* For the PEC-PV tandem system, two hematite photoanodes connected back-to-back are placed within the cell and held in place by a stainless-steel connector, which also serves for establishing electrical contact, and a Si PV module is attached to the back window behind the photoanodes. Two charged NiOOH AEs are placed within the cell, on each side of the photoanodes, and perpendicular to the photoanodes, also held in place by stainless-steel connectors. The lid has an inlet that connects to a water reservoir for electrolyte level control and an optional inlet for temperature measurement.

*Figure 4: Electrochemical hydrogen generation cell.* Four discharged $Ni(OH)_2$ AEs and one platinized titanium mesh cathode are held by stainless-steel connectors that are inserted through the lid. The lid has an inlet for purge gas and a gas outlet.

*Figure 5: Auxiliary electrode swapping scheme.* In (A) a reserve stock of four charged AEs (Y = 4) is placed at the solar site, out of which two AEs (X = 2) are placed inside the oxygen cell for the first cycle. Similarly, four discharged AEs are placed in the hydrogen cell, out of which only two AEs are connected to the oxygen cell. In (B) and (C), after the first cycle the two (now discharged) AEs in the oxygen cell (yellow highlight) are replaced with two new (charged) AEs from the reserve stock, while the two (now charged) AEs in the hydrogen cell (red highlight) are disconnected, and the other pair is connected instead (C). In (D), after the second cycle the AEs are swapped between the cells, after all the AEs in the solar (oxygen) site have been discharged (yellow highlight) and all the AEs in the hydrogen cell have been charged (red highlight).

*Figure 6: Current – potential (voltage) characteristic curves of hematite photoanodes and a silicon photovoltaic module.* (A) I-E LSV curve of two hematite photoanode samples, and (B) I-V curves of a photovoltaic module, both scanned at 20 $mV·s^{-1}$. In (A) the two photoanodes were scanned separately (blue and red curves) and when connected back-to-back (black curves) under dark (dashed line curves) and 1-Sun illumination (solid line curves). Sample II was also scanned in back illumination (dotted line blue curve). In (B) the PV module was scanned alone (red curve), and behind the oxygen cell with and without the photoanode stack (blue and

*green curves, respectively). The I-E curve of the back-to-back photoanode stack is overlaid in (B), black curve.*

***Figure 7: Current – potential (voltage) characteristic curves of the individual separate-cell system components.** (A) I-E LSV scans of the platinized Ti mesh cathode and the hematite photoanode stack with and without illumination. Each electrode was connected separately as the working electrode in a three-electrode configuration with an Hg/HgO/1 M NaOH reference electrode and a nickel foil counter electrode. The dotted green and blue lines show the OCPs of the charged and discharged AEs, respectively, measured in the same configuration. (B) I-V curves measured in two-electrode configuration for the hydrogen cell and the oxygen cell with and without illumination. (C) I-V curves of the PV module and of the entire separate-cell PEC system with and without illumination. (✗) presents the current and voltage measured throughout the ten cycles carried out with simulated sunlight (Figure 8). The I-E and I-V curves were obtained at a scan rate of 20 mV·s$^{-1}$.*

***Figure 8: Indoor separate-cell PEC water splitting demonstration.** (A) Short-circuit current ($I_{sc}$) traces measured for the separate-cell PEC system throughout ten water splitting cycles under solar simulated illumination without any additional bias during water splitting. (B) The voltage measured between the photoanode stack and the cathode (solid line black curve), as well as the potentials of the charging and discharging AEs relative to reference electrodes in their respective cells (red and blue curves). The potential difference between the AEs (green curve) is subtracted from the measured voltage (solid line black curve) to show the expected voltage of an equivalent single cell system (dashed line curve).*

***Figure 9: Outdoors separate-cell PEC water splitting demonstration.** The separate-cell PEC system was placed outdoors (A) and its short-circuit current was measured under natural sunlight for 8 hours, between 10:20 AM and 6:20 PM (B).*

***Figure 10: Separate-cell PEC water splitting with simultaneous hydrogen and electricity generation.** (A) The I-V characteristic curves of a single PV module having an area of 50 cm$^2$, as well as two PV modules having a total area of 100 cm$^2$ connected in series and in parallel. The I-V curves for the separate-cell PEC system and the 25 Ω load are added to give the I-V curve of the separate-cell system and load in a parallel connection; the I-V characteristic curves of the combined system is represented by the dotted blue line. (B) The current of the combined system and of its individual components, i.e., the load and the separate-cell PEC system. The measured combined system voltage is represented by the red line. The inset shows the system's electronic scheme.*

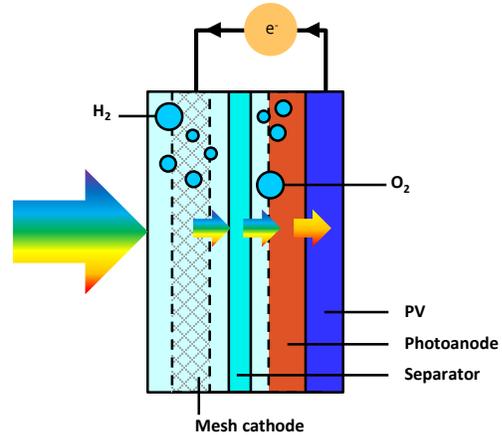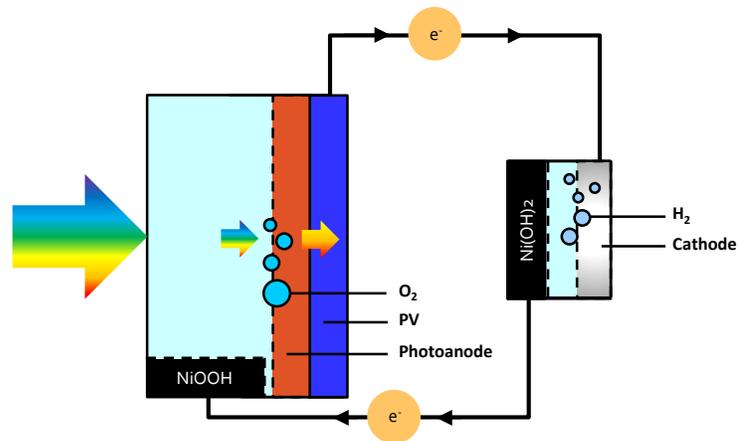

Figure 1

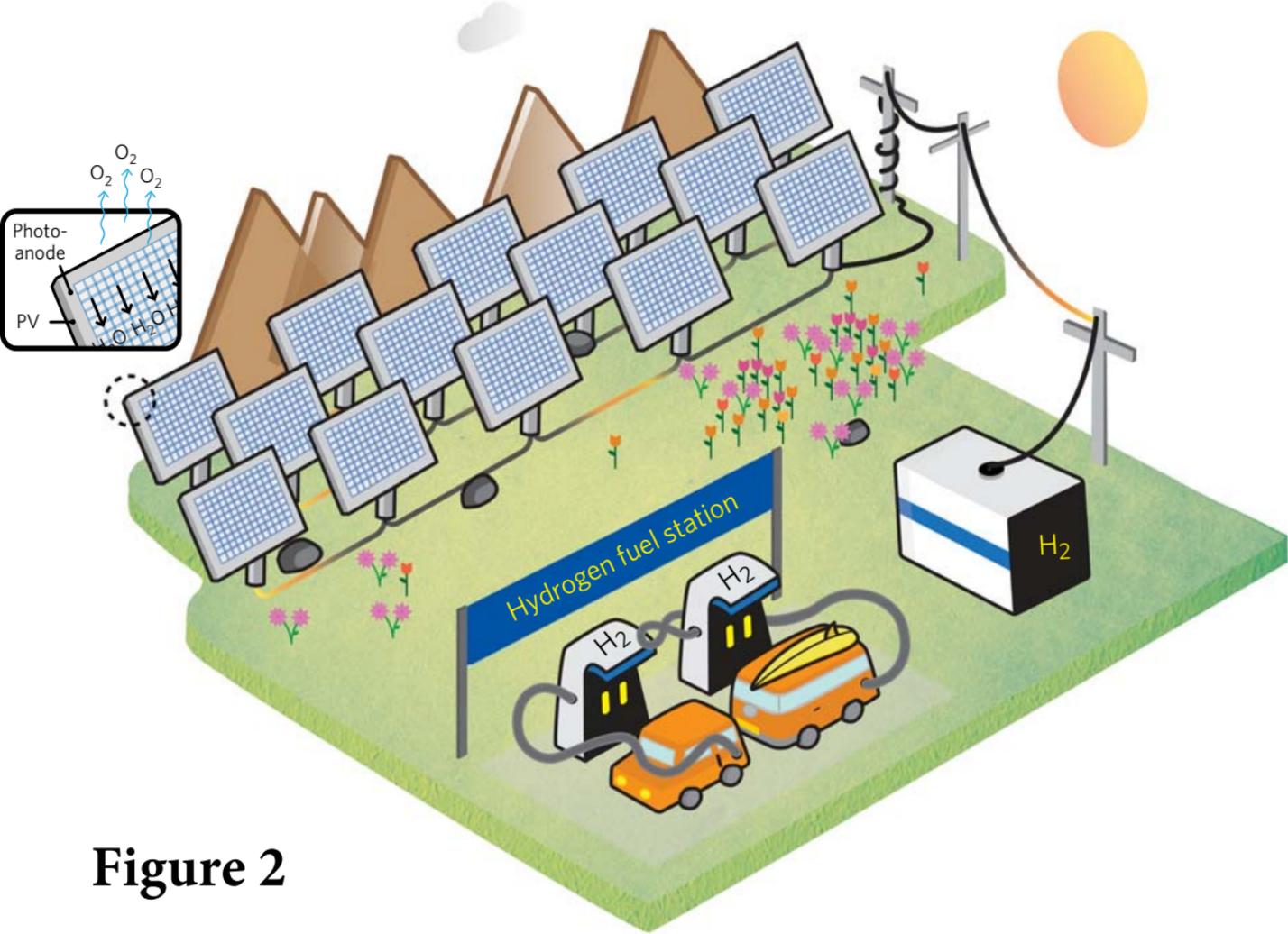

**Figure 2**

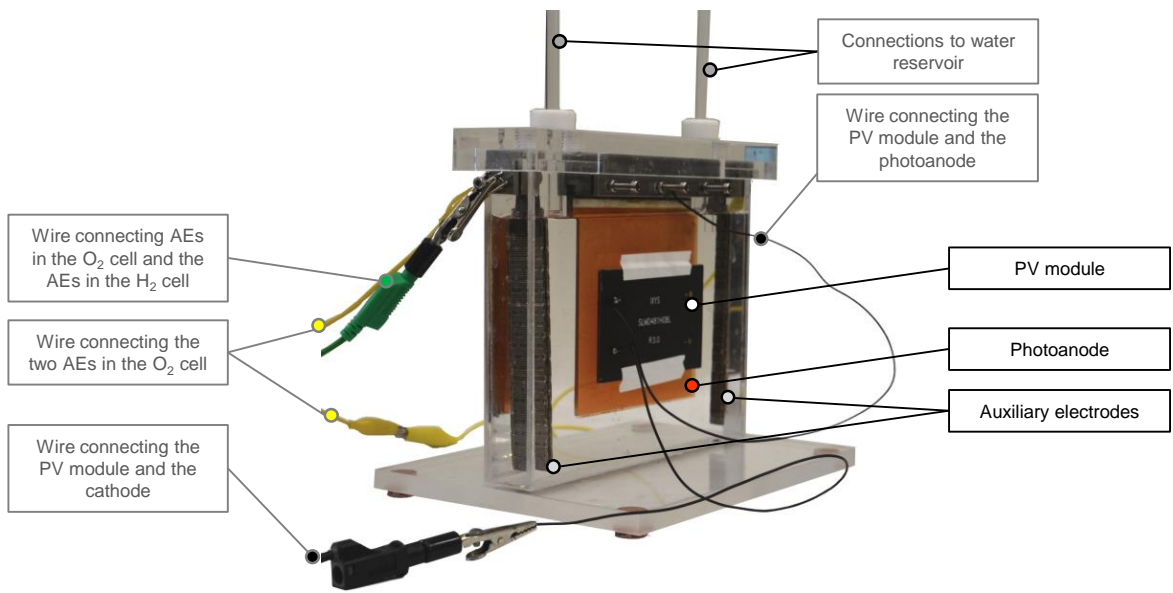

**Figure 3**

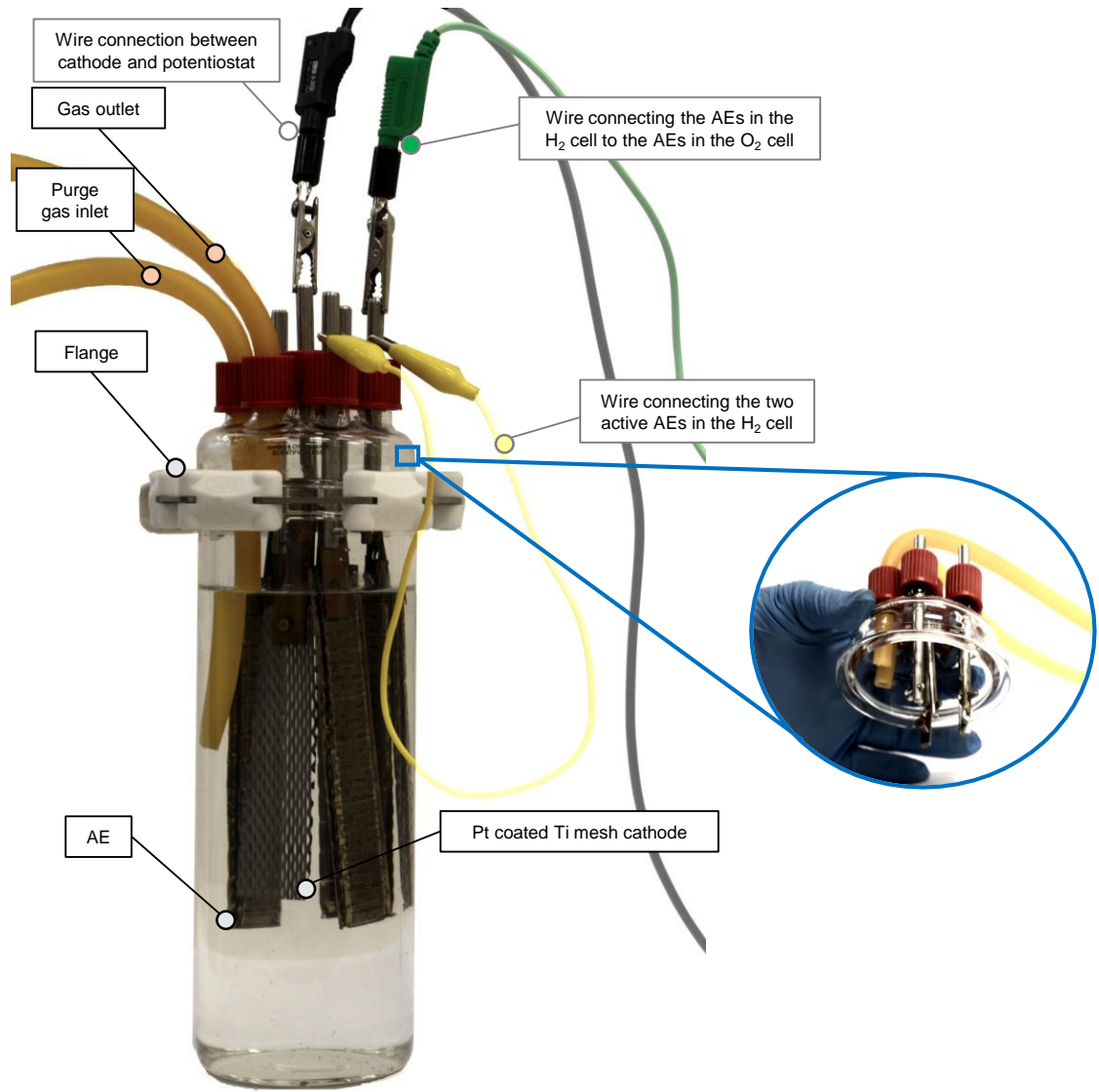

**Figure 4**

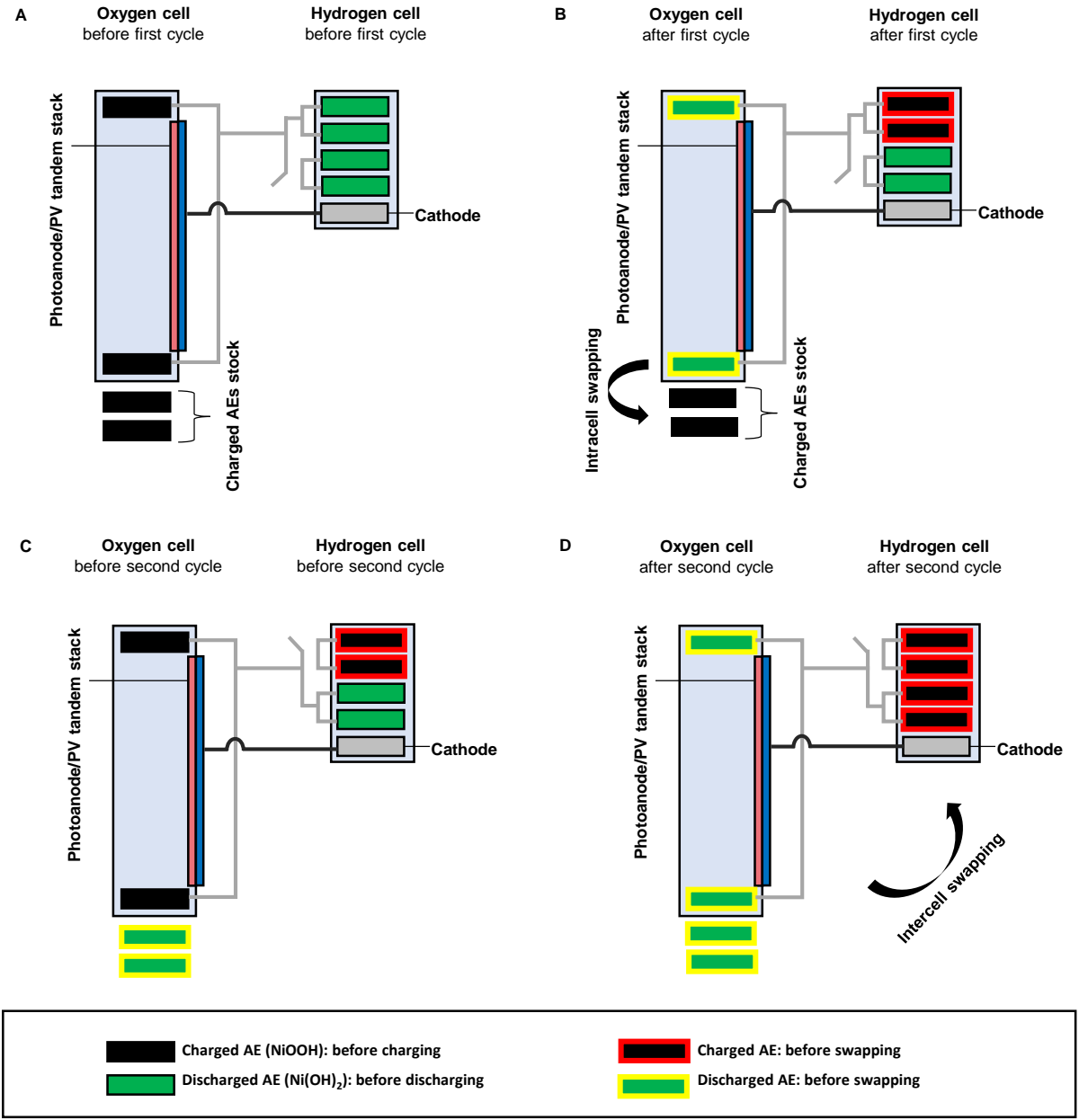

**Figure 5**

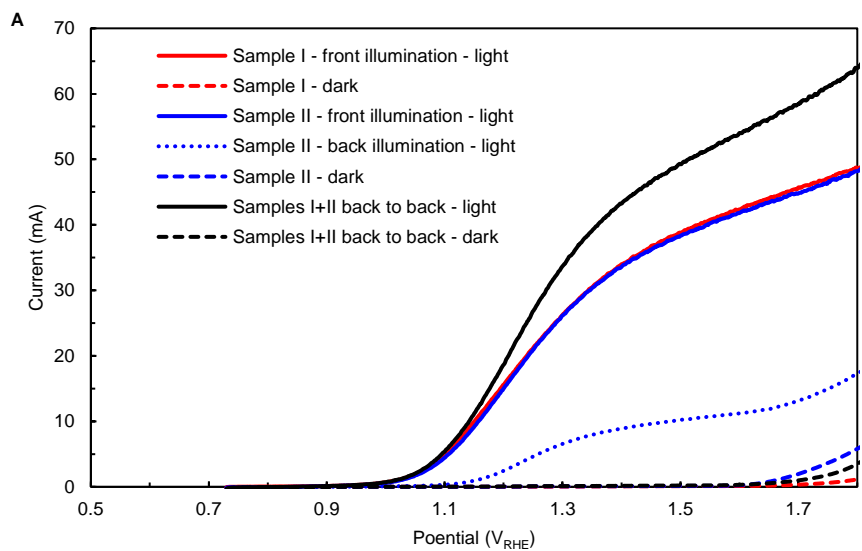
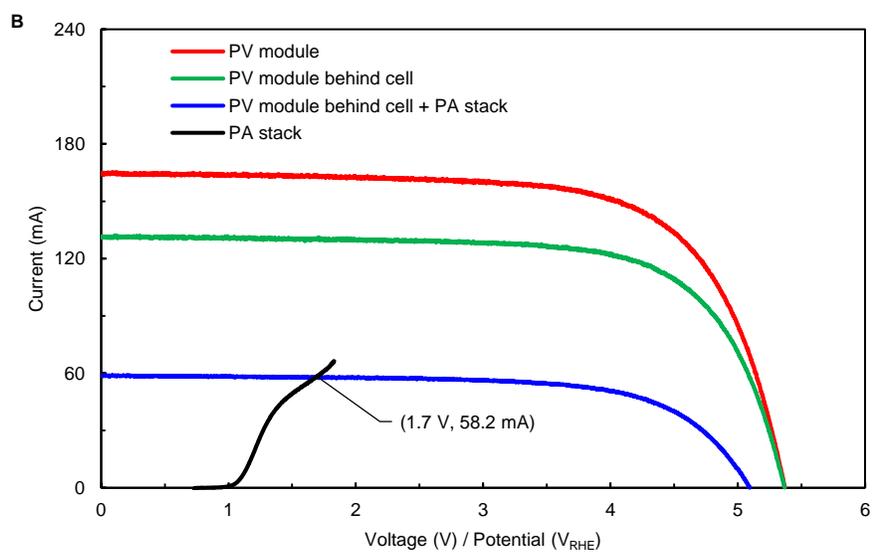

Figure 6

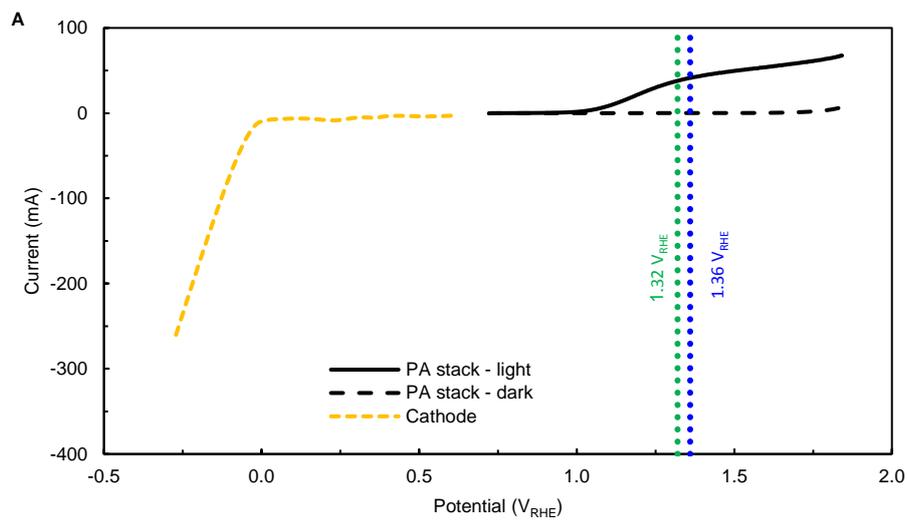

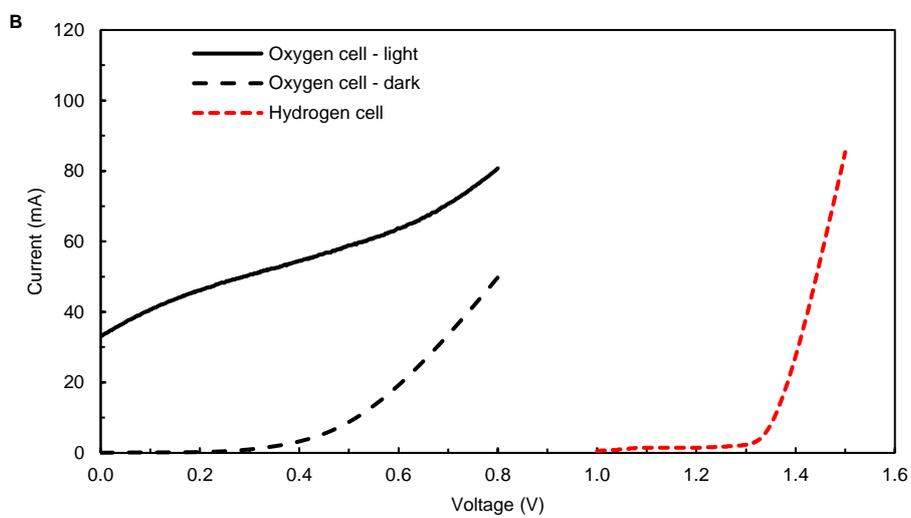

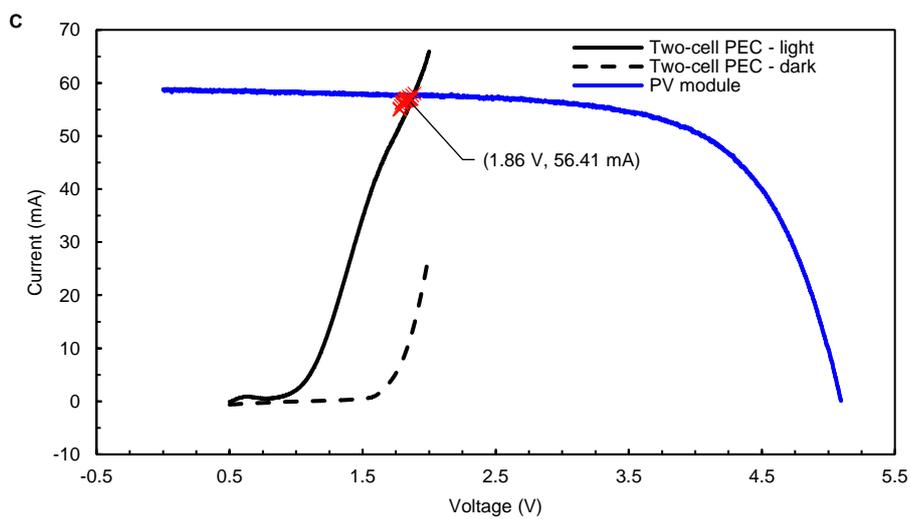

Figure 7

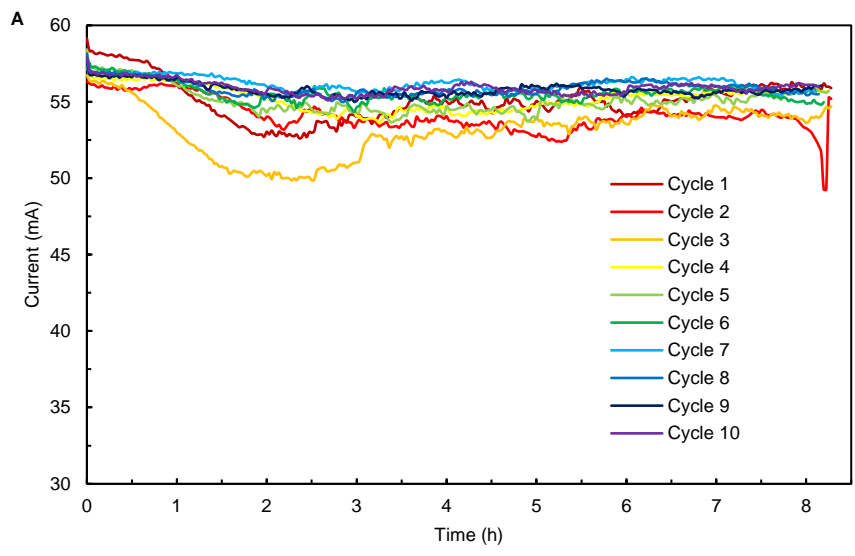
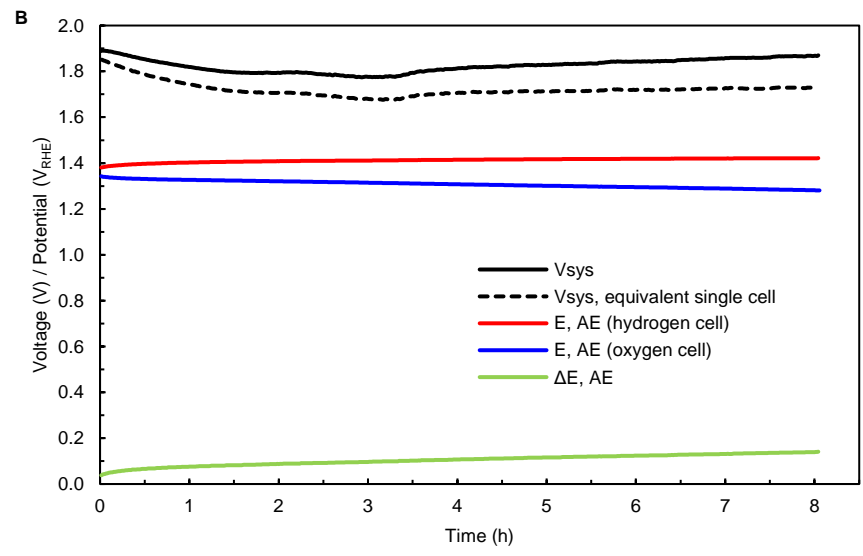

Figure 8

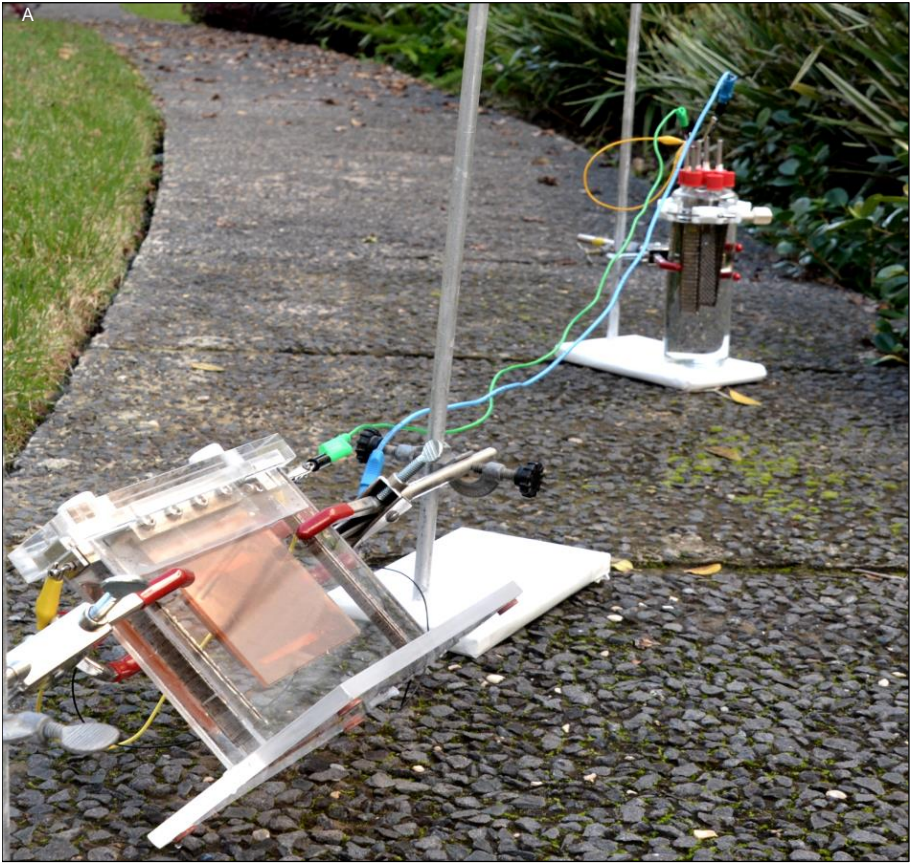
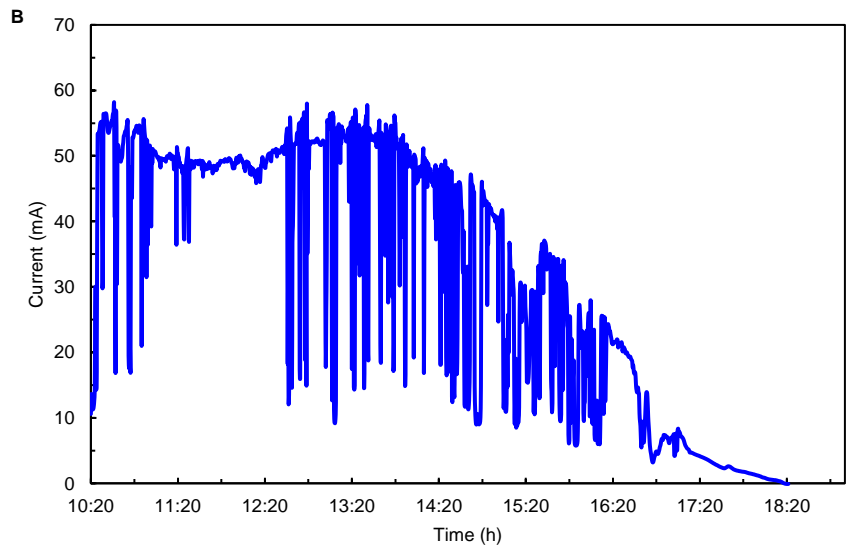

**Figure 9**

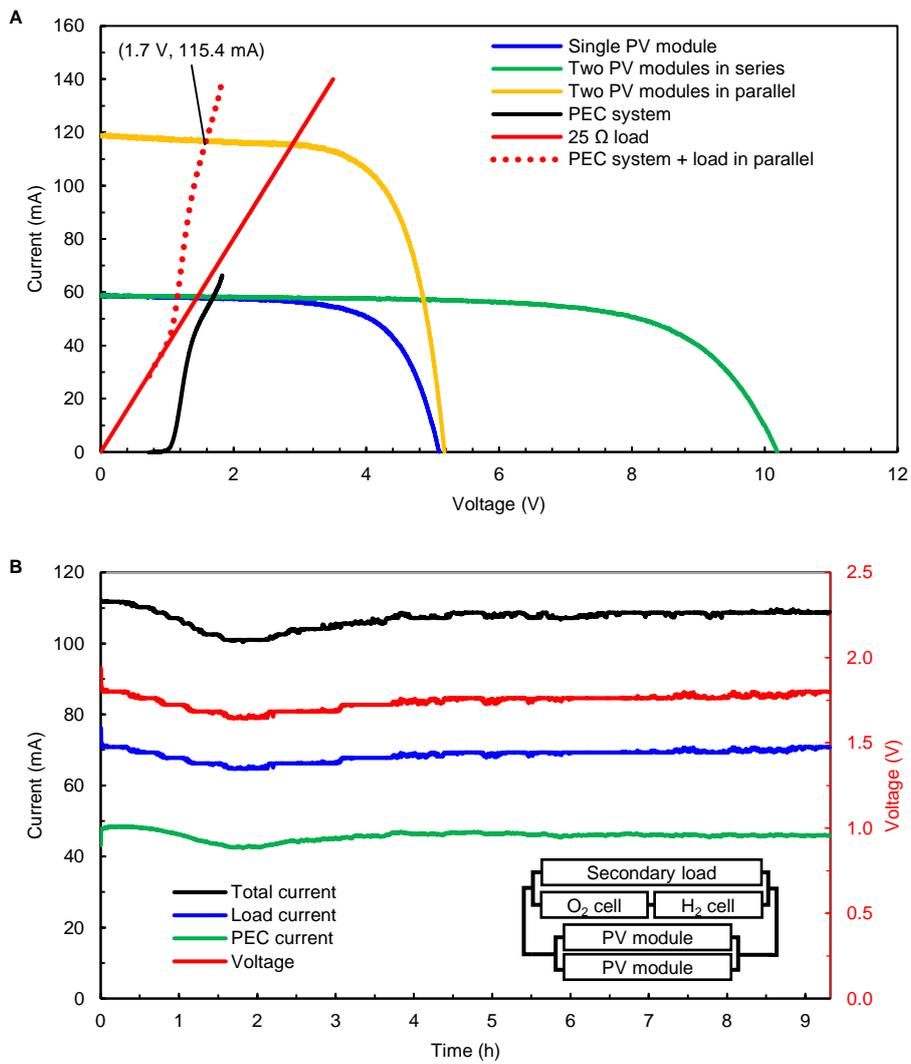

Figure 10